\documentclass[12pt]{article}
\newtheorem{theorem}{Theorem}
\newtheorem{lemma}{Lemma}
\newtheorem{prop}[lemma]{Proposition}

\begin{document}
\bibliographystyle{plain}

\begin{titlepage}

\title{Central limit theorems for nonlinear hierarchical sequences of 
        random variables}

\author{Jung M. Woo and Jan Wehr \\
        University of Arizona }

\date{ July 10, 1999 }

\maketitle

\begin{abstract}
We study central limit theorems for 
certain nonlinear sequences of random variables. In particular, we prove
the central limit theorems for the bounded conductivity 
of the random resistor networks on hierarchical lattices
\end{abstract}


Running head:

\end{titlepage}

\section{Introduction}
Let $D$ be a closed interval of ${\bf R}$ (bounded or unbounded) and
$f:D^k \rightarrow D\,\, (k\ge 2)$ be a continuous function.
Let $x_0$ be a $D$-valued random variable 
with a distribution $\mu_0$.
We define a hierarchical sequence of random variables 
$\{x_n : n=0,1,2, \dots \}$ inductively by
\begin{equation}
x_n = f(x_{n-1,1},x_{n-1,2},\dots, x_{n-1,k})
\end{equation}
where for each $n\ge 1$ $x_{n-1,1} \, x_{n-1,2} \, \dots, x_{n-1,k} $ 
are independent identically distributed (shortly IID) random variables
having the same distribution as $x_{n-1}$.
For each $n \ge 0$ let us denote a distribution of $x_n$ by $\mu_n$.
We thus write
\begin{eqnarray}
x_n = \underbrace{ f\circ f \circ \cdots \circ f}_{ n\, \mbox{times} } (x_0)
\nonumber \\
\mu_n =\underbrace{ f\circ f\circ \cdots \circ f}_{n\, \mbox{ times } } (\mu_0)
\end{eqnarray}
We are interested in central limit theorem for the hierarchical sequence $x_n$ 
under suitable assumptions of $f$.
Assume that $f$ is twice continuously differentiable 
at $(c_n, c_n, \dots, c_n)$ where $c_n =E[x_n]$.
Let
$$
\alpha_{n,i} = \frac{\partial f}{\partial u_i} (c_n,c_n,\dots, c_n)
$$

By expanding $x_{n+1}$ at $(c_n,c_n,\dots,c_n)$, we get
\begin{equation}
x_{n+1} = f(c_n,c_n,\dots,c_n) + \sum_{i=1}^{k} \alpha_{n,i} ( x_{n,i} - c_n)
         +\mbox{ second order } + \cdots
\end{equation}
For each $n$ let
$$
d_n = f(c_n,c_n,\dots,c_n) - c_n \sum_{i=1}^{k} \alpha_{n,i}
$$ 
and
\begin{equation}
z_n = x_{n+1} - \sum_{i=1}^{k} \alpha_{n,i} x_{n,i}
\end{equation}
Then
$$
z_n = d_n +
     \mbox{ second order } + \cdots
$$
Hence $x_{n+1}$ can be viewed as
a linear function of $x_{n,i}$ plus 
some random variable $z_n$ of ``small'' variance.
The following central limit theorem applies to sequences of this kind 
(not necessarily defined by iterations).

\begin{theorem}
For all $n\ge 0$, let $k_n$ be a positive integer greater than or equal to $2$.
Assume that a sequence of real-valued random variables
$\{x_{n} , n=0,1,2,\cdots \}$ satisfies the following recursive relation.
\begin{equation}
x_{n+1}  =\sum_{i=1}^{k_n} \alpha_{n,i} x_{n,i}  + z_n
\end{equation}

where for each $n\ge 0$ $\alpha_{n,i} ( i \le k_n) $ are real numbers,
$z_n$ is a real-valued random variable with $E[z_n^2] < \infty$,
and $\{ x_{n,i}: i=1,\cdots,k_n\}$ are
IID random variables with same distribution as $x_{n}$.
We also assume that $E[x_0^2] < \infty$ and that there exists a $\delta>0$  
such that for all $n$
$ |\alpha_{n,i} | \ge \delta $ and $ | \alpha_{n,j} | \ge \delta$ for at least
two distinct indices $i$ and $j$ (``mixing'' property).
Let $\lambda_n = \sqrt{ \sum_{i=1}^{k_n} \alpha_{n,i}^2 }$ and
$\sup_n \lambda_n < \infty$.
Furthermore, assume that there exist  $\delta_1>0$, $\delta_2>0$, $C_1>0$ 
and $C_2>0$ with $\delta_1<\delta_2$ such that
for any $n$  
\begin{eqnarray}
Var[x_n] \ge C_1^2 \lambda_0^2 \lambda_1^2 \dots \lambda_{n-1}^2
                ( 1 - \delta_1 )^{2n} \nonumber \\
Var[z_{n}]\le C_2^2 \lambda_0^2\lambda_1^2 \lambda_2^2 \dots
                \lambda_{n-1}^2 (1- \delta_2)^{2n} \label{varbds}
\end{eqnarray}
Then $\frac{x_{n} -E x_{n} }{ \sqrt{Var[x_n]} } $ 
converges in distribution to a unit normal variable.
\end{theorem}
We will prove Theorem 1 in section 2. 
To apply Theorem 1 to some hierarchical sequence 
of real-valued random variables $x_n$, 
we have to prove the variance bounds (6).
In order to prove the variance bounds in many cases,
we need weak laws of large numbers and large deviation estimates.
In section 3, we will prove weak laws of large numbers and 
large deviation estimates for some class of functions $f$.

We will say that a continuous function 
$f: D^k \rightarrow D$ (where $D$ is a closed interval) 
is averaging if the following three conditions hold:
\\
1. For all $u_i\in D$ ( $i=1,2,\cdots,k$)
\begin{equation}
\min_i u_i\le f(u_1,u_2,\dots,u_k) \le \max_i u_i 
\end{equation}
2. $f$ is monotone, that is, 
for all $u_i$ and $u_i'$ ( with $u_i\le u_i'$ )
\begin{equation}
f(u_1,u_2,\dots,u_k) \le f(u_1',u_2',\dots,u_k')
\end{equation}
3. if equality holds in 2 then 
\begin{equation}
\mbox{ $u_i=u_i'$ and $u_j =u_j'$ 
for at least two distinct indices $i$ and $j$}
\end{equation}
The following Theorem 2, proved in section 4,
 is one of the applications of Theorem 1.

\begin{theorem}
Let $[a,b]$ be a closed and bounded interval with $a<b$.
Let $f:[a,b]^k \rightarrow [a,b]$ be averaging.
Assume that the essential range 
of $[a,b]$-valued random variable $x_0$
$$
{\mathcal{R} }(x_0) = \{ u \in [a,b]: 
  \mbox{ for any $\epsilon>0$  } {\bf P}[ |x_0 -u | > \epsilon ] >0 \}
$$
is connected and consists of more than
one point.
Suppose there exists  $c\in (a,b)$ such that 
$x_n$ converges to $c$ in probability
(a weak law of large numbers ).
Also assume that there exists $\epsilon_1>0$ such that 
$(c-\epsilon_1, c + \epsilon_1) \subset (a,b)$ and
$f$ is twice continuously differentiable on
$ (c -\epsilon_1, c + \epsilon_1)^k$ and that
$\frac{\partial f}{\partial u_i} (c,c,\dots,c) >0$ 
for two distinct indices $i$.
Then
$$
\frac{x_n - E[x_n]}{ \sqrt{ Var[x_n]} }
$$
converges to a unit normal variable in distribution.
\end{theorem}
Another application of Theorem 1 is to prove
a central limit theorem for the conductivity of random resistor networks
on hierarchical lattices. 
Let us define hierarchical lattices first.
A detail description of various hierarchical lattices
can be found in~\cite{GK} and~\cite{ScS}.
Let ${\bf G} = ( {\mathcal{S} }, {\mathcal{B} }, (s_t,s_b) )$ be a graph with 
site set $\mathcal{S}$, bond set $\mathcal{B}$, top site $s_t$ and bottom site
$s_b$. The top site and bottom site are called surface sites since
they are the sites where we apply the potential difference. All other sites are
called internal sites. Let $k$ be the total number of bonds of ${\bf G}$.
We define a hierarchical lattice 
${\bf H} =\{ {\bf H_n } : n=0,1,2\}$ (constructed from
the fixed graph ${\bf G}$) inductively.
At level $0$, ${\ H_0} $ is a single bond.
At level $n$, a graph ${\bf H_n}$ is constructed from ${\bf H_{n-1} }$
by replacing each bond of ${\bf H_{n-1} }$ by the fixed graph ${\bf G}$.
Since at each level each bond
is replaced by the graph ${\bf G}$
having $k$ bonds, ${\bf H_n}$ consists of total $k^n$ bonds, 
many internal sites and 
two surface sites where we apply the potential difference. 
We will assume that the graph ${\bf G}$ used in 
constructing the hierarchical lattice 
${\bf H_n}$, is connected and there exist at least two
bond-disjoint self-avoiding paths connecting two surface sites of ${\bf G}$. 
we will also assume that for any bond $b$ of ${\bf G}$ there exists a 
self-avoiding path passing through the bond $b$ 
and connecting two surface sites.
Furthermore we will assume any self-avoiding path connecting 
two surface sites of ${\bf G}$ has at least length $2$.
Let us order all bonds of ${\bf G}$ in a specific way.
For each $i=1,\dots,k$ let 
$u_i\ge 0$ be the conductance of the $i$-th bond of ${\bf G}$. 
Let $h(u_1,u_2,\dots,u_k)$ be the conductance of the graph ${\bf G}$.
Define $f: [0,\infty)^k \rightarrow [0, \infty)$ by
\begin{equation}
f(u_1,u_2,\dots,u_k) = \frac{h(u_1,u_2,\dots,u_k)}{h(1,1,\dots,1)}
\end{equation}
Then this function $f$ is nonnegative continuous, monotone increasing,
 homogeneous of degree one, concave, and $f(1,1,\dots,1)=1$
(See~\cite{Shn}). 
Let $x_{0,i}$ be IID random variables with a common distribution $\mu_0$
where $x_{0.i}$ represents the random conductance of $i$-th bond of ${\bf G}$.
Then the random conductance $\widehat{x}_n$ of the hierarchical lattice
${\bf H_n}$ is given by
\begin{equation}
\widehat{x}_n = 
   \underbrace{h \circ h \circ \cdots \circ h}_{n \, \mbox{times} } (x_0)
\end{equation}
Clearly a normalized random conductance 
$x_n = h(1,1,\dots,1)^{-n} \widehat{x}_n$ is
\begin{equation}
x_n =\underbrace{f \circ f \circ \cdots \circ f}_{n \, \mbox{times} } (x_0)
\end{equation}
Let $x_0$ be a bounded nonnegative random variable.
Let
$$
p ={\bf P} [ x_0 >0]
$$ 
and 
$$
g(p) = {\bf P} [ x_1 >0]
$$ 
Note that the function $g(p)$ 
does not depend on the distribution of $x_0$ 
but depends only on the parameter $p$.
In~\cite{Shn}, Shneiberg proved that there exists $c(\mu_0)$ such that
the normalized conductivity $x_n$ converges to $c( \mu_0)$ in probability
and that there exists a unique fixed point 
$p_c({\bf G} ) \in (0,1)$ of $g$ (that is $g(p) =p$) such that
\begin{equation}
c(\mu_0) = \left\{ \begin{array}{ll}
                  0 & \mbox{if  $\,p\le p_c$} \\
                 \mbox{positive} & \mbox{if  $\,p > p_c$}
                \end{array}
           \right.
\end{equation}
The following Theorem 3 is a central limit theorem for the conductivity
of random resistor networks on hierarchical lattices. 
The proof will be given in section 4.
\begin{theorem}
Let $\widehat{x_n}$ be the conductivity of a random resistor network on a 
hierarchical lattice ${\bf H}$ 
constructed using a fixed graph ${\bf G}$ described above. 
Assume that a random variable $x_0$ is
nonnegative, bounded and not almost surely a constant.
Also let ${\bf P}[x_0>0]> p_c({\bf G})$.
where $p_c({\bf G})$ is a unique fixed point of $g$ mentioned above.
Then the random variables
$$
\frac{\widehat{x}_n - {\bf E}[\widehat{x}_n]}
{ \sqrt{ {\bf Var} [\widehat{x}_n]} }
$$
converge in distribution to a unit normal variable.
\end{theorem}
Remark: For ${\bf P}[x_0>0] \le p_c({\bf G})$ 
we expect a non-Gaussian behavior for $\widehat{x}_n$.
\section{Proof of Theorem 1}
In this section we use the notations introduced in Theorem 1 as well as 
the following.
For all $n\ge 0$ let 
\begin{equation}
y_n = \sum_{i=1}^{k_n} \alpha_{n,i} x_{n,i}
\end{equation}
and for all $n \ge 1$ let
\begin{eqnarray}
\widetilde{x_n} = 
\frac{x_n - {\bf E} [ x_n]}{\lambda_0\lambda_1\cdots\lambda_{n-1} } 
\nonumber \\
\widetilde{y_n} = 
\frac{y_n -{\bf E} [y_n] }{\lambda_0\lambda_1\cdots\lambda_{n-1} } \\
\widetilde{z_n} =
\frac{z_n -{\bf E} [z_n] }{\lambda_0\lambda_1\cdots\lambda_{n-1} } \nonumber  
\end{eqnarray}

Then the assumptions in Theorem 1 say that for all $n$
\begin{eqnarray}
\widetilde{x_{n+1}} = 
 \frac{1}{\lambda_n} ( \widetilde{y_n} + \widetilde{z_n}) \nonumber \\
\sqrt{ {\bf Var} [ \widetilde{ y_n} ]} = \lambda_n
     \sqrt{  {\bf Var}[ \widetilde{ x_n} ] } \nonumber \\
\sqrt{ {\bf Var} [ \widetilde{ x_n} ] } 
  \ge C_1 ( 1 - \delta_1)^n \label{eq12}\\
\sqrt{ {\bf Var} [ \widetilde{ z_n} ] } \le C_2 ( 1 - \delta_2)^n \nonumber 
\end{eqnarray}

\begin{lemma}
Let
$$
\sigma_{\infty} = \lim_{n \rightarrow \infty}
\sqrt{ {\bf Var} [ \widetilde{x_n} ] }
$$
Then $\sigma_{\infty}$ exists and is a positive real number.
Moreover there exist positive real constants $C_3$, $C_4$, and $C_5$ such that
for all $n$
\begin{eqnarray}
\inf_{n} \sqrt{ {\bf Var}[ \widetilde{ x_n} ]} \ge C_3 \nonumber \\
\sup_{n} \sqrt{ {\bf Var}[ \widetilde{ x_n} ]} \le C_4 \label{eq13} \\
\sup_{n} \sqrt{ {\bf Var}[ \widetilde{ y_n} ]} \le C_5 \nonumber
\end{eqnarray}
\end{lemma}

\bigskip
Proof: 
{From} (16)
and by the triangle inequality,
\begin{eqnarray*}
& &\left|\sqrt{ {\bf Var}[ \widetilde{x_{n+1}}]}   - \sqrt{ {\bf Var}
  [ \widetilde{ x_n} ] } 
   \right| \\
& & = \left|\sqrt{ {\bf Var} [ \widetilde{x_{n+1}}]}   - \frac{1}{\lambda_n} 
             \sqrt{ {\bf Var}[ \widetilde{ y_n} ] } 
      \right|   \\
& &  \le \frac{1}{\lambda_n} \sqrt{ {\bf Var}[ \widetilde{z_n}] }
  \le  \frac{1}{\underline{\lambda} }  C_2 ( 1- \delta_2)^{n}
\end{eqnarray*}
where $\underline{\lambda} = \inf_{n} \lambda_n >\delta$.
By summing up the above inequality for all $n\ge m$, we can find a positive 
constant $C_6$ such that for all $m$
$$
\sum_{n=m}^{\infty} \left| \sqrt{ {\bf Var}[ \widetilde{x_{n+1} } ] }
    - \sqrt{ {\bf Var}[ \widetilde{x_n}]} 
          \right| \le C_6 ( 1 - \delta_2 )^m
$$
Since the above series converges absolutely, in particular 
$\sigma_{\infty}$ exists (and is finite) and for all $m$
$$
\left| \sigma_{\infty} - \sqrt{ {\bf Var}[ \widetilde{ x_m} ] } 
 \right| 
  \le C_6 (1 - \delta_2)^m
$$
Therefore from (16)
and from the above inequality, for all $m$ we have
\begin{equation}
C_1 ( 1 -\delta_1)^m \le \sqrt{ {\bf Var}[ \widetilde{ x_m} ] } 
\le \sigma_{\infty} + C_6 (1 - \delta_2)^m \label{eq131}
\end{equation}
Since the above inequality (18) is true for all $m$,
from $\delta_2 > \delta_1$
$\sigma_{\infty}$ cannot be zero.
Three inequalities of Lemma 1
follow immediately from the positivity 
of $\sigma_{\infty}$, and (16).

\bigskip


\begin{lemma}
Let $C_7$ be any positive constant, and let
$X$ and $Y$ be random variables with zero means 
and variances less than $C_7^2$.
Then for all $t\,\, ( |t| < \frac{1}{C_7} )$
\begin{equation}
\left| \ln {\bf E} [\exp (itX)] - \ln {\bf E}[\exp (itY )] \right|
 \le 4C_7 t^2 \sqrt{ {\bf E}[(X - Y)^2]} \label{eq14}
\end{equation}
Also we have
\begin{equation}
\lim_{t \rightarrow 0} \frac{ \ln {\bf E}[ \exp ( itX) ]} {t^2} 
  = -\frac{1}{2} {\bf Var}[X] \label{eq15}
\end{equation}
\end{lemma}
Proof:
Since $| e^{iu} -1 - iu| \le \frac{u^2}{2}$ for any real number $u$,
$$
 \left| {\bf E}[ \exp (itX) ] - 1 \right| 
  \le t^2 \frac{ {\bf E} [X^2] }{2}
$$
Therefore for $|t| < \frac{1}{C_7}$,
the value ${\bf E}[ \exp (itX) ]$ lies inside of a circle of radius 1/2 
centered at $(1,0)$
in the complex plane (similarly to $Y$).
Hence for $|t| < \frac{1}{C_7}$
the logarithmic function $\ln {\bf E}[ \exp (itX) ]$ is single-valued.
Since $|\ln u - \ln v| \le 2 |u -v|$ for any complex number $u$ and $v$
with $|u -1 |\le \frac{1}{2}$ and $|v-1 | \le \frac{1}{2}$, 
for $|t|< \frac{1}{C_7}$ we have
\begin{eqnarray*}
\lefteqn{ \left| \ln  {\bf E}[ \exp (itX)]  - 
     \ln  {\bf E} [\exp (itY)] \right| } \nonumber \\
  & \le & 2 \left| {\bf E}[ \exp (itX)]  - {\bf E} [\exp (itY)] 
          \right|
     =  2 \left| {\bf E}[\,\, \exp (itX) 
                          (\, 1 - \exp (it(Y-X))\, )\,\,] 
          \right|      \nonumber \\
  & \le & 2 {\bf E} [\,\, | \exp (itX) -1 | | 1- \exp (it(Y-X)) |\,\, ]  + 
       2  {\bf E} [\, | 1- \exp (it(Y-X)) | \,]  \nonumber\\
  & \le & 2t^2 {\bf E} [|X||Y-X|] + t^2 {\bf E} [(Y-X)^2]   \nonumber  \\
  & \le & 4t^2 C_7  \sqrt{ {\bf E}[(X-Y)^2] }  
\end{eqnarray*}
where in the last part we used 
Cauchy-Schwarz inequality and the triangle inequality.
The second part of Lemma 2 is elementary and the proof can be found in many
basic probability textbooks (for example see~\cite{Dur}).

For all $n$ let us define a characteristic function $\phi_n$ of 
the normalized random variable $\widetilde{x_n}$ by
$$
\phi_n (t) = {\bf E}[ \exp ( it \widetilde{x_n} ) ]
$$

\begin{lemma}
There exists a positive constant $C_8$ such that 
for any  $|t| <\frac{1}{C_8}$
$$
\lim_{n \rightarrow \infty} \ln \phi_{n}(t) 
= - \frac{1}{2} \sigma_{\infty}^2 t^2
$$ 
(the convergence is uniform on $|t|< \frac{1}{C_8}$ ).
Therefore
$\widetilde{x_n}$ converges in distribution 
to a normal variable with mean zero and
variance
 $\sigma_{\infty}^2$.
\end{lemma}
Proof:
{From} Lemma 1 and Lemma 2, it follows that there exists a positive constant
$C_8$ such that for any $|t|<\frac{1}{C_8}$ and
any $n$
$$
\left| \ln {\bf E}[ \exp ( it\widetilde{x_{n+1}}) ]  - 
 \ln {\bf E}[ \exp (it \frac{1}{\lambda_n} \widetilde{y_n} ) ]
\right| \le 4 C_8 t^2 
\sqrt{ {\bf Var}\left[ \frac{\widetilde{z_n}}{\lambda_n} \right] }
$$
For all $n$ and $i\le k_n$,
let $\beta_{n,i} = \frac{\alpha_{n,i}}{\lambda_n}$.
Then {from} the above inequality and from (16), 
there is a positive constant $C_9$ such that
for all $|t| \le \frac{1}{C_8}$
\begin{equation}
\left| \ln \phi_{n+1}(t) - \sum_{i=1}^{k_n} \ln \phi_n 
     (\beta_{n,i} t)
\right| \le C_9 t^2 ( 1- \delta_2)^n       \label{char1}
\end{equation}
Note that for any real $\gamma_{n,j}$ satisfying $\sum_{j} \gamma_{n,j}^2 =1$, 
we have
\begin{equation}
\sum_{j} \left| \ln \phi_{n+1}(\gamma_{n,j} t) - \sum_{i=1}^{k_n} \ln \phi_n 
     (\gamma_{n,j} \beta_{n,i} t)
       \right| \le C_9 t^2 ( 1- \delta_2)^n       \label{char2}
\end{equation}
{From} the above inequality (22), it follows that 
for all $|t| \le \frac{1}{C_8}$
\begin{eqnarray}
& & \left| \ln \phi_{n+m} (t) - \sum_{i_1,i_2,\dots,i_m }
  \ln \phi_n ( t \beta_{n,i_1} \beta_{n+1,i_2} \dots \beta_{n+m-1,i_m} ) 
    \right| 
\nonumber \\
& & \le  C_9 t^2 \sum_{p=0}^{m-1} ( 1- \delta_2)^{n+p}
\le \frac{C_9}{\delta_2} t^2 ( 1- \delta_2)^n \label{char3}
\end{eqnarray}
Note that by one of the assumptions in Theorem 1 (``mixing'' property)
$$
\max_{i_1,i_2,\dots,i_m}
   \beta_{n,i_1}\beta_{n+1,i_2} \dots \beta_{n+m-1,i_m} 
$$
converges to zero when $m \rightarrow \infty$.
{From} the second part of Lemma 2, it follows that
for any $n$ and any $\epsilon_1>0$ 
there exists an $M_1(n,\epsilon_1)$ such that
for all $m\ge M_1$
\begin{eqnarray*}
& &\left| \ln \phi_n 
      ( t \beta_{n,i_1} \beta_{n+1,i_2} \cdots \beta_{n+m-1,i_m} )
 + \frac{1}{2} t^2 
( \beta_{n,i_1}\beta_{n+1,i_2}\dots \beta_{n+m-1,i_m} )^2
  {\bf Var}[\widetilde{x_n} ] \right|  \\
& &\le \epsilon_1 t^2 
( \beta_{n,i_1}\beta_{n+1,i_2}\dots \beta_{n+m-1,i_m} )^2
\end{eqnarray*}
Hence for any $m\ge M_1$, 
by summing the above inequality over all $i_1, \, i_2, \dots, i_m$, we obtain
\begin{equation}
\left| \sum_{i_1,i_2,\dots, i_m}
  \ln \phi_n ( t \beta_{n,i_1}\beta_{n+1,i_2}\dots \beta_{n+m-1,i_m} )
 + \frac{1}{2} t^2 {\bf Var}[ \widetilde{x_n} ] \right|
\le \epsilon_1 t^2  \label{char4}
\end{equation}
{From} (23) and (24),
using the triangle inequality, we have for all $n$, $\epsilon_1>0$,
and $|t| < \frac{1}{C_8}$
$$
\limsup_{m\rightarrow\infty}
 \left|  \ln \phi_{n+m}(t) + \frac{1}{2} t^2 {\bf Var}[ \widetilde{x_n} ]
 \right| 
\le \epsilon_1 t^2 +  
   \frac{C_9}{\delta_2} t^2 (1 -\delta_2)^n 
$$
Since $\epsilon_1>0$ is arbitrary, the proof is finished by taking the 
limit $n \to \infty$. \\

\underline{Proof of Theorem 1}.
{From} Lemma 3, 
$\frac{ \widetilde{x_n} }{ \sqrt{ {\bf Var}[ \widetilde{x_n} ]} }$
converges to a unit normal variable in distribution.
Since 
$$
\frac{x_n - {\bf E}[x_n] }{ \sqrt{ {\bf Var}[ x_n] } } 
= \frac{ \widetilde{x_n} }{ \sqrt{ {\bf Var}[ \widetilde{x_n} ]} }
$$
we finish the proof of Theorem 1.

\section{Law of large numbers}

In this section we state and prove  weak law of large numbers and large 
deviation estimate used later to prove central limit theorems for hierarchical
sequences of real-valued random variables.
To be more precise, we define some terminology below.
A sequence of real-valued random variables $x_n$ 
satisfies a weak law of large numbers if there exists a 
real number $c$ such that for any $\epsilon>0$
\begin{equation}
\lim_{n \rightarrow \infty} {\bf P}[ | x_n -c | > \epsilon] =0
\end{equation}
We say that $x_n$ satisfies a large deviation estimate if
there exists a real number $c$ such that 
for any $\epsilon_1>0$ and $\epsilon_2>0$, 
there exists $M>0$ such that
for all $n$
\begin{equation}
{\bf P}[ | x_n - c| > \epsilon_1 ] \le M \epsilon_2^n
\end{equation}
The above is not the strongest large deviation estimate satisfied by
hierarchical sequences studied in this paper, but
it is a sufficient condition for central limit theorems.

\begin{prop}
Let $D$ be a closed interval (bounded or unbounded) of ${\bf R}$.
Let $f:D^k \rightarrow D$ be differentiable and 
$x_0$ be a $D$-valued random variable 
with ${\bf E}[ \exp ( \delta |x_0| ) ] < \infty$ for some $\delta>0$.
Also assume that there exists $\epsilon>0$ such that
$$
\sup_{(u_1, \dots, u_k)} \sum_{i=1}^{k} | 
\frac{\partial f}{ \partial u_i} (u_1,u_2,\dots,u_k))|
< 1 -\epsilon
$$ 
Then a weak law of large numbers and a large deviation estimate hold.
\end{prop}
Proof:
Let us define $g(u) = f(u,u,\dots,u)$.
Then by the mean value theorem $| g(u) - g(v) | \le ( 1 - \epsilon) | u -v|$.
Thus $g$ is a contraction mapping on $D$ and, consequently has a unique fixed 
point $c\in D$, that is $f(c,c,\dots,c) =c$.
Then by the mean value theorem and our assumptions on $f$,
$$
|x_{n+1} - c | = | f(x_{n,1}, x_{n,2},\dots,x_{n,k}) - f(c,c,\dots,c) |
\le (1- \epsilon )\max_{i\in \{1,\dots,k\} } | x_{n,i} - c |
$$
Therefore for any positive real number $v$
$$
{\bf P} [ |x_{n+1} -c | >v] \le k {\bf P} 
  [ | x_{n} -c | > \frac{v}{1 - \epsilon} ]
$$
{From } the above inequality,
for any $v>0$ and for all $n$ we have
$$
{\bf P} [ |x_n -c| >v] \le k^n 
 {\bf P} [ |x_0 -c| >  \frac{v}{ (1 - \epsilon)^n} ]
 \le k^n {\bf E} [ \exp ( \delta ( | x_0 -c| - \frac{v}{ (1 -\epsilon)^n} ) ) ]
$$
where in the last equation we used Chebyshev's inequality.
Weak law of large numbers and large deviation estimate 
follows immediately from the above inequality.

\begin{prop}
Let $[a,b]$ be a closed and bounded interval of ${\bf R}$ with $a< b$,
and $f:[a,b]^k \rightarrow [a,b]$ be averaging.
Also let $x_0$ be a $[a,b]$-valued random variable.
Furthermore assume that there exists $c\in [a,b]$ such that
$x_n$ converges to $c$ in probability (weak law of large numbers).
Then a large deviation estimate holds for the same constant $c$.
\end{prop}

Proof:
By the symmetry of definitions,
it is enough to show that for any $\epsilon >0$ and for any $u <c$  
there exists $M>0$ such that for all $n$
\begin{equation}
{\bf P}[ x_n < u ] < M \epsilon^n
\end{equation}
In case $c =a$, the above inequality is obvious, hence we will assume that
$c \in (a, b]$.
Let $F_{n}(u) = {\bf P}[ x_n \le u]$ be the distribution function of
$x_n$  and let
\begin{equation}
Q=\{ u \in [a,b] : \mbox{for any } \epsilon > 0 \mbox{ there exists an } M 
 \mbox{ such that } F_{n}(u) \le M \epsilon^{n} \}
\label{eq:ldvset}
\end{equation}
We only need to prove that $\sup Q = c$. 
Without loss of generality we will assume that $f$ is symmetric under 
the exchange of $k$ variables.
First we will show that for some positive $\epsilon_1$, $a + \epsilon_1 \in Q.$
Since $f(a,c,c,\dots,c) > a$ (by monotonicity)
there exists $\epsilon_1\in (0, c - a)$ such that
$$
f(a, c-\epsilon_1,c-\epsilon_1,\dots,c-\epsilon_1) > a + \epsilon_1
$$

For such $\epsilon_1$
\begin{eqnarray*}
& & {\bf P}[x_{n+1} \le a + \epsilon_1] \\
& & \le k! \, {\bf P}[ f(x_{n,1},x_{n,2},\dots,x_{n,k}) \le  a + \epsilon_1;
              x_{n,1}\le x_{n,2} \le \dots \le x_{n,k} ;
              x_{n,1} \le a + \epsilon_1 ]  \\
& &  \le k!\, {\bf P}[ x_{n,1} \le  a + \epsilon_1; 
           f(a,x_{n,2},x_{n,2},\dots, x_{n,2}) \le a + \epsilon_1 ]  \\
& & \le k!\,{\bf P}[ x_{n,1}\le a + \epsilon_1;  x_{n,2}\le c - \epsilon_1]\\
& & \le k! \, {\bf P}[x_n \le c - \epsilon_1] {\bf P}[x_n \le a + \epsilon_1]
\end{eqnarray*}
{From} the above inequality and by the weak law of large numbers,
it follows that $a + \epsilon_1 \in Q$.
Now we will prove that $\sup Q = c$.
We will prove this by contradiction. Let $u = \sup Q < c$.
First note that $u \ge a + \epsilon_1$.
Since $f(u,c,c,\dots,c) > u$ (by monotonicity)
there exists $\epsilon_2 \in (0, \epsilon_1)$ such that
$$
f(u-\epsilon_2, c-\epsilon_2,c-\epsilon_2,\dots,c-\epsilon_2) > u + \epsilon_2
$$

For such $\epsilon_2$
\begin{eqnarray*}
& & {\bf P}[x_{n+1} \le u + \epsilon_2] \\
& & \le k! \, {\bf P}[ f(x_{n,1},x_{n,2},\dots,x_{n,k}) \le  u + \epsilon_2;
              x_{n,1}\le x_{n,2} \le \dots \le x_{n,k} ;
              x_{n,1} \le u + \epsilon_2 ]  \\
& & \le k! \, {\bf P}[x_{n,1} \le u - \epsilon_2 ] +  \\
& &  k! \, {\bf P}[ u - \epsilon_2 \le x_{n,1} \le u + \epsilon_2;
   f(u - \epsilon_2, x_{n,2},x_{n,2},\dots, x_{n,2}) \le u + \epsilon_2] \\
& & \le k! \, {\bf P}[x_{n,1} \le u -\epsilon_2] + 
  k! \, {\bf P} [ x_{n,1} \le u + \epsilon_2;  x_{n,2} \le c - \epsilon_2 ] \\
& & \le k! \, {\bf P} [ x_n \le u - \epsilon_2] +
     k!\, {\bf P}[x_n < c - \epsilon_2 ] {\bf P}[ x_n \le u + \epsilon_2]
\end{eqnarray*}
Take any $\epsilon>0$. By weak law of large numbers and
{from}  $ u - \epsilon_2 \in Q$ (by the assumption of $u$ ),
there exist $N$ and $M>0$ such that for all $n\ge N$
$$
{\bf P}[x_{n+1} \le u + \epsilon_2] \le M (\frac{\epsilon}{2})^n + 
\frac{\epsilon}{2} {\bf P}[ x_n \le u + \epsilon_2]
$$
Obviously by induction that there exists $K>0$ such that 
for all $n$ $F_n(u + \epsilon_2) \le K \epsilon^n$,
which is a contradiction by definition of $Q$.
Hence we finished the proof of Proposition 5.

\section{Variance bounds}
In this section we will prove Theorem 2 and Theorem 3.
Let $[a,b]$ be a closed and bounded interval of ${\bf R}$.
Throughout this section, we assume that 
$f: [a,b]^k \rightarrow [a,b]$ is averaging and $x_0$ is $[a,b]$-valued 
random variable. We also assume that there exists a real number 
$c\in (a,b)$ such that $x_n$, defined as in (1),
converges to $c$ in probability. 
In section 3, we showed for such a function $f$, our sequence
$x_n$ satisfies a large deviation estimate.
Furthermore We assume that there exists $\epsilon_1>0$ such that 
$f$ is twice continuously differentiable on 
$(c - \epsilon_1, c+ \epsilon_1)^k$.
We use the same notations introduced in section 1, section 2 and the following 
as well.
Let
$$
\alpha_i = \frac{\partial f}{\partial u_i}(c,c,\dots,c)
$$
$$
\lambda = \sqrt{ \sum_{i=1}^{k} \alpha_i^2 }
$$
$$
\underline{\lambda} = \inf_n \lambda_n
$$
Let a random variable $x_0'$ be an independent copy of $x_0$, and for all $n$
define $x_n'$, $y_n'$ and $z_n'$ similarly 
as we defined $x_n$, $y_n$, and $z_n$ in (2),(4) and (14).
We have the following variance bounds.
\begin{prop}
For any $\epsilon>0$ there exists $M>0$ such that for 
all $n$
$$
{\bf Var}[z_n] \le M( \lambda + \epsilon)^{4n}
$$
\end{prop}

Proof:
Since $x_n$, taking values in a bounded interval $[a,b]$,
converges to $c$ in probability, $c_n$ converges to $c$.
Note that there exists $\epsilon_1>0$ such that 
$f$ is twice continuously differentiable on 
$(c -\epsilon_1, c+ \epsilon_1)^k$. 
Hence for such $\epsilon_1>0$ 
there exists $N_1$ such that for all $n \ge N_1$
$c - \epsilon_1 < c_n < c + \epsilon_1$.
Therefore there exists $M_1>0$ such that for all $n \ge N_1$
\begin{equation}
{\bf Var}[ z_n] \le M_1{\bf E}[ ( x_n -c_n)^4]
\end{equation}
Since for any $\epsilon_2 >0$ and for all $n\ge N_1$
$$
{\bf Var}[z_n] \le M_1 \epsilon_2^2 
{\bf E}[ (x_n -c_n)^2; |x_n -c_n| \le \epsilon_2 ]
       + M_1 {\bf E}[ (x_n -c_n )^4; | x_n -c_n| \ge \epsilon_2 ],
$$
applying a large deviation estimate to the above inequality,
for any $\epsilon_3$ 
there exists $N_2>N_1$ such that
for all $n\ge N_2$
$$
{\bf Var}[z_n]
 \le \epsilon_3^2  {\bf Var}[x_n] + \epsilon_3^{2n}
$$

{From} the triangle inequality and the above inequality, for
all $n \ge N_2$ we have 
$$
\sqrt{ {\bf Var}[ x_{n+1} ]} 
\le \lambda_n \sqrt{ {\bf Var}[ x_n ]} 
 + \sqrt{ {\bf Var}[ z_n ]} 
\le ( \lambda_n + \epsilon_3)
  \sqrt{ {\bf Var}[ x_n ]} + \epsilon_3^n
$$
Hence, by the induction on $n$, for any $\epsilon_3>0$ 
there exists $M_2>0$ such that
 for all $n$  
\begin{equation}
{\bf Var}[x_n] \le M_2 (\lambda +  2 \epsilon_3 )^{2n} 
\end{equation}
Also note that
$$
{\bf E}[ (x_n - x_n')^4 ]  = 
2 {\bf E}[ (x_n -c_n)^4 ] + 6 ( {\bf E}[ (x_n -c_n)^2] )^2
$$
Therefore from (29)
there exists $M_3>0$ and $N_1>0$ such that for all $n\ge N_1$
\begin{equation}
{\bf Var}[z_n] \le M_3 {\bf E}[ (x_n - x_n')^4 ]
\end{equation}
{From} (14)
\begin{eqnarray*}
{\bf E}[ (y_n -y_n')^4 ] & \le & 
(\sum_{i=1}^{k} \alpha_{n,i}^4) {\bf E}[ (x_n - x_n')^4] 
  + \lambda_n^4 ( {\bf E}[ (x_n -x_n')^2] )^2  \\
  & \le & \lambda_n^4 \left( {\bf E}[ (x_n -x_n')^4 ] +
      ({\bf E}[ (x_n -x_n')^2] )^2  \right)
\end{eqnarray*}
Also since there exists $M_4>0$ such that $(a+ b)^4 \le M_4 (a^4  + b^4)$ 
for all real values $a$ and $b$, and by (4) 
there exists $M_5>0$ such that
$$
{\bf E}[ (z_n -z_n')^4 ] \le M_4
( {\bf E}[ (z_n -d_n)^4 ] + {\bf E}[ (z_n' - d_n)^4 ] )
 \le M_5 {\bf E}[ (x_n -c_n )^8]
$$

Since for any $\epsilon_4>0$ there exists $M_6$ such that
for all $a$ and $b$ $(a+b)^4 \le (1 + \epsilon_4) a^4 + M_6 b^4$,
for such $\epsilon_4$ there exists $M_7$ such that
\begin{eqnarray}
& & {\bf E}[(x_{n+1} -x_{n+1}')^4]  = 
{\bf E}[ ( y_n - y_n' + z_n - z_n' )^4 ] \nonumber \\
        &  & \le (1+ \epsilon_4) {\bf E}[ (y_n -y_n')^4 ] + 
             M_6 {\bf E}[ (z_{n} - z_n')^4]  \\
        &  & \le (1+ \epsilon_4)
     \lambda_n^4 \left( {\bf E}[ (x_n -x_n')^4 ]  + 
           ( {\bf E}[ (x_n -x_n')^2] )^2  \right)
            +  M_7 {\bf E} [( x_n - c_n)^8]  \nonumber 
\end{eqnarray}
Since, by a large deviation estimate, we can show that
for any $\epsilon_5>0$ there exists $N_3>N_2$ such that for all
$n\ge N_3$
$$
M_7 E[ (x_n -c_n)^8 ] \le \epsilon_5 E[ (x_n -c_n)^4 ] + 
                       \epsilon_5^n
$$
{From} the above two inequalities, for any $\epsilon>0$ there exists an
$N_3>0$ such that for all $n\ge N_3$
$$
{\bf E}[ (x_{n+1} - x_{n+1}')^4 ] \le (\lambda + \epsilon)^4 ( 
        {\bf E}[ (x_n -x_n')^4 ] + 
         {\bf E}[ (x_n -x_n')^2] )^2  )
      + \epsilon^n
$$
Hence from (29), (30) and  the above inequality, by induction, we prove the 
desired statement of Proposition 6.

\begin{prop}
For all $\epsilon>0$ let
\begin{equation}
C(n,\epsilon) =
\sqrt{ {\bf E}[ (x_n - x_n')^2; 
 |x_n -c| \le \epsilon ; |x_n' - c | \le \epsilon ] }
\end{equation}
Suppose that for any $\epsilon>0$ there exists $N>0$ such that
$C(N,\epsilon) >0$. 
Then for any $\epsilon>0$ there exists
$\epsilon_1>0$, $M_1>0$ and $N_1>0$
such that for all $n\ge N_1$
\begin{equation}
C(n,\epsilon_1) > M_1 ( \lambda - \epsilon )^n
\end{equation}
\end{prop}

Proof:
By the averaging property of $f$ and by mean value theorem,
for any $\epsilon>0$
there exists $\epsilon_1>0$ such that
for all
$|x_{n,i} - c | \le \epsilon_1$ and $|x_{n,i}' -c | \le \epsilon_1$
where $i=1,2,\dots,k$, we have
$$
|x_{n+1} -c| \le \epsilon_1 \mbox{ and } | x_{n+1}' -c | \le \epsilon_1
$$
and
$$
| x_{n+1} - x_{n+1}' - \sum_{i=1}^k 
\alpha_i  (x_{n,i} -x_{n,i}')  |
\le \epsilon \sum_{i=1}^k | x_{n,i} - x_{n,i}' |
$$
By applying the triangle inequality to the above inequality,
\begin{eqnarray*}
& & C(n+1, \epsilon_1) \ge 
\sqrt{ {\bf E}[ (x_{n+1} -x_{n+1}' )^2; |x_{n,i} -c| \le \epsilon_1; 
  |x_{n,i}' -c| \le \epsilon_1 |,
   \mbox{i=1,2,\dots,k} ] }   \\
& &  \ge \sqrt{ {\bf E}[ (\sum_{i=1}^k \alpha_i (x_{n,i} -x_{n,i}') )^2;
  | x_{n,i} -c| \le \epsilon_1, |x_{n,i}' -c|\le \epsilon_1,
   \mbox{i=1,2,\dots,k} ] }  \\
& & -
\epsilon \sqrt{ {\bf E}[ ( \sum_{i=1}^k | (x_{n,i} -x_{n,i}') | )^2 ;
  | x_{n,i} -c| \le \epsilon_1, |x_{n,i}' -c|\le \epsilon_1,
   \mbox{i=1,2,\dots,k} ] }
\end{eqnarray*}
Hence using the independence and the triangle inequality, 
we obtain
\begin{equation}
 C(n+1, \epsilon_1) 
   \ge \lambda {\bf P}[ |x_n -c| \le \epsilon_1]^{k-1}
     C(n,\epsilon_1)
  - \epsilon k  C(n, \epsilon_1) 
\end{equation}

By weak law of large numbers, there exists $N>0$ such that
for all $n\ge N$,
$$
C(n+1, \epsilon_1) \ge 
     ( \lambda ( 1 - \epsilon) - k \epsilon )
      C(n, \epsilon_1)
$$
Since the above is true for any $\epsilon>0$ 
($\epsilon_1$ and $N$ depending on $\epsilon$ ),
Proposition 7 follows immediately by induction.

\underline{Proof of Theorem 2}:
Suppose $x_n$ and $f$ satisfy all assumptions of Theorem 2.
We only need to prove that the same sequence $x_n$ satisfies all assumptions
of Theorem 1. We only need to check (6).
Since $f$ is averaging, $\alpha_i \ge 0$ (by monotonicity ) and
$\sum_{i=1}^{k} \alpha_i =1$. 
Also $\alpha_i>0$ at least for two distinct indices 
(by assumptions of Theorem 2). Hence 
$$
0< \lambda= \sqrt{ \sum_{i=1}^{k}\alpha_i^2 } < 1
$$
For all $n$, define the essential range of $x_n$ by
$$
{\mathcal{R}}(x_n) = \{ u \in [a,b] : \mbox{ for any $\epsilon>0$  
${\bf P}[|x_n - c| > \epsilon ] >0$ }\}
$$
Since ${\mathcal{R}}(x_0)$ is connected and consists of more than one point,
by assumptions of Theorem 2 and
by averaging property of $f$ 
${\mathcal{R}}(x_n) ={\mathcal{R}}(x_0)$ is connected 
and consists of more than one point. In particular for all
$n$ and $\epsilon_1>0$ $C(n,\epsilon_1)>0$. Since $0<\lambda<1$,
by Proposition 6 and 
Proposition 7, the condition (6) follows immediately.

\underline{Proof of Theorem 3}:
Since 
$$
\frac{\widehat{x}_n - {\bf E}[\widehat{x}_n]}{\sqrt{ {\bf Var}[\widehat{x}_n]}}
=
\frac{x_n - {\bf E}[x_n]}{\sqrt{ {\bf Var}[x_n]}},
$$
it is enough to prove the central limit theorem 
for $x_n$.
Note that $f$ is averaging and since ${\bf P}[x_0>0] >p_c({\bf G})$,
there exists $c>0$ such that $x_n$ converges to $c$ 
in probability (see~\cite{Shn}).
Moreover $f$ is at least twice continuously differentiable 
in the neighborhood of $(c,c,\dots,c)$. Since each self-avoiding path 
connecting two surface sites of ${\bf G}$ 
has length at least $2$,
there exists $\delta>0$ such that
$\frac{\partial f}{\partial u_i} (c,c,\dots,c) >\delta$ 
at least two distinct indices $i$.
Hence we only need to prove (6) to in order apply Theorem 1 for $x_n$.
Let $s_t$ and $s_b$ be the top and bottom sites of ${\bf G}$ respectively.
For each bond $i$ of ${\bf G}$, 
let $s_1(i)$ and $s_2(i)$ be the two sites of bond $i$
The variational principle of conductivity says that
for any conductance $w_i\ge 0$ of $i$-th bond of ${\bf G}$, we have
\begin{equation}
f(w_1,w_2,\dots,w_k) = 
C \min_{v} \sum_{i=1}^{k} w_i \left( v(s_1(i)) - v(s_2(i)) \right)^2
\end{equation}
where minimum is taken over all real-valued functions $v$,
defined on site set of ${\bf G}$,
satisfying $v(s_t)=1$ and $v(s_b)=0$ and 
the normalization constant $C>0$ is chosen to satisfy $f(1,1,\dots,1)=1$
(see~\cite{DS} for variational principle of conductivity).
Let $A$ and $B$ be two bond-disjoint self-avoiding paths connecting
two surface sites of ${\bf G}$.
Let 
$$
u_i = \left\{ \begin{array}{ll}
               1 & \mbox{if $i\in A$}  \\
               0 & \mbox{otherwise}
              \end{array}
      \right.
$$

$$
u_i' = \left\{ \begin{array}{ll}
               1 & \mbox{if $i\in B$}  \\
               0 & \mbox{otherwise}
              \end{array}
      \right.
$$

$$
\eta_A = f(u_1,u_2,\dots,u_k)
$$
$$
\eta_B = f(u_1',u_2',\dots,u_k')
$$
Then by (36) $\eta_A>0$ and $\eta_B >0$.
Let $b>a$.
For each bond $i$ of ${\bf G}$, let
$$
{w_i} = \left\{  \begin{array}{ll}
                            b  & \mbox{if $i\in A$} \\
                            a  & \mbox{otherwise}
                           \end{array}
                  \right.
$$    
By applying (36) to $w$, we get
\begin{equation}
a + (b-a) \eta_A 
\le f({w_1},{w_2},\dots,{w_k}) 
\le b - (b-a) \eta_B
\end{equation}
Take any $\epsilon>0$.
Since $x_0$ is not almost surely a constant, by averaging property of $f$
$x_n$ is not almost surely a constant. Also by weak law of large numbers, 
there exists $N$ such that
$$
{\bf P}[ | x_N -c | \le \frac{\epsilon}{2}] >0
$$
For such $N$,
we can choose two distinct real numbers $a_N\in{\mathcal{R}}(x_N)$ and 
$b_N\in{\mathcal{R}}(x_N)$ such that
$|a_N -c| < \frac{\epsilon}{2}$
Without loss of generality, let us assume that
$a_N < b_N$.
For all $n > N$, define $a_n$ and $b_n$ inductively by
$$
a_n = f(a_{n-1},a_{n-1},\dots,a_{n-1}) = a_{n-1}
$$
and
$$
b_n = f(w_{n-1,1},w_{n-1,2},\dots,w_{n-1,k})
$$
where
$$
w_{n-1,i} = \left\{ \begin{array}{ll}
                  b_{n-1} & \mbox{if $i\in A$} \\
                  a_{n-1} & \mbox{otherwise}
                  \end{array}
           \right.
$$
Clearly for all $n\ge N$, 
$a_n$ and $b_n$ are distinct and elements of ${\mathcal{R}}(x_n)$.
Furthermore by (37) $b_n -a_n$ converges to zero when $n$ goes to infinity.
Hence for large $n$, $C(n, \epsilon)>0$.
Since $0<\lambda <1$,
(6) follows immediately from Proposition 6 and Proposition 7.

\end{document}